\documentclass[aps,pra,reprint,nofootinbib,superscriptaddress,twocolumn,showpacs,showkeys,longbibliography,amsmath,amssymb]{revtex4-2}
\usepackage{graphicx}
\usepackage{amsmath}
\usepackage{dcolumn}
\usepackage{bm}
\usepackage{braket}
\usepackage{subfigure}
\usepackage[colorlinks,bookmarks=false,hypertexnames=false,citecolor=blue,linkcolor=red,urlcolor=blue]{hyperref}
\usepackage[english]{babel}
\usepackage{multirow}
\usepackage{booktabs}
\usepackage{float}
\usepackage{array} 
\begin{document}
\preprint{APS/123-QED}	
	
\title{Gapless Cornwall--Jackiw--Tomboulis description of a weakly interacting one-dimensional Bose gas}
\author{Yi Zhang}
\affiliation{Department of Physics, Zhejiang Normal University, Jinhua 321004, People's Republic of China}
\author{Xiaoran Ye}
\affiliation{Department of Physics, Zhejiang Normal University, Jinhua 321004, People's Republic of China}
\author{Ziheng Zhou}
\affiliation{Department of Physics, Zhejiang Normal University, Jinhua 321004, People's Republic of China}
\author{Zhaoxin Liang}\email[Corresponding author:~] {zhxliang@zjnu.edu.cn}
\affiliation{Department of Physics, Zhejiang Normal University, Jinhua 321004, People's Republic of China}

\date{\today}	

\begin{abstract}
The weakly interacting one-dimensional (1D) Bose gas is a paradigmatic system for studying quantum fluctuations, and its equation of state (EOS) provides a stringent test of many-body theory. Conventional Bogoliubov theory accurately describes the EOS of a weakly interacting three-dimensional (3D) Bose gas. In 1D, long-wavelength phase fluctuations destroy true condensate order even at zero temperature, invalidating the conventional expansion about a condensate. Here we develop a gapless Cornwall–Jackiw–Tomboulis (CJT) effective field theory for the EOS of a weakly interacting 1D Bose gas. Combining the background, Dyson, and number equations eliminates the infrared-singular transverse correlator from the bulk closure. The resulting quadratic equation determines the chemical potential and energy density, both of which recover the negative 1D Lee–Huang–Yang (LHY) correction. At relative order \(\gamma_{\rm LL}\), the energy coefficient falls \(1/6\) below the exact Lieb–Liniger (LL) value. With a common physical infrared cutoff, the quantum depletion is nonnegative and increases logarithmically with the size of the coherent region within the small-depletion regime.
\end{abstract}
\maketitle

\section{Introduction}\label{sec:intro}
Ultracold atomic gases provide a controllable setting for studying
how dimensionality and interactions shape quantum many-body
behavior~\cite{Bloch2008,Rigol2011}. In one dimension (1D),
long-wavelength phase fluctuations remain important even at weak
coupling and shape the low-energy hydrodynamics~\cite{Rigol2011}.
The equation of state (EOS) connects microscopic interactions to
the chemical potential, compressibility, and thermodynamic sound
velocity. Fluctuation effects on dilute quantum gases are also
evident in studies of dipolar droplets, density patterns, and
droplets in atomic
mixtures~\cite{FerrierBarbut2016,Chomaz2016,Wenzel2017,Semeghini2018,Cabrera2018,DErrico2019}.
In 1D, measurements of the EOS of individual
gases~\cite{SalcesCarcoba2018} and hydrodynamic calculations based
on the EOS~\cite{Kopycinski2022} connect thermodynamics to
observable properties. Measurements of Tan's contact in the
Lieb--Liniger (LL) gas probe short-range
correlations~\cite{Huang2025}. Local pair correlations provide
proposed routes to the free energy and
entropy~\cite{KerrFreeEnergy2024,Watson2024}. Studies of one-body
correlations and momentum distributions also characterize coherence
in homogeneous and trapped gases~\cite{DeRosi2023,Takacs2025}.

For repulsive contact interactions, the LL model gives the exact
ground-state EOS through its Bethe-ansatz solution~\cite{Lieb1963}.
The harmonic-fluid description captures its low-energy
behavior~\cite{Haldane1981PRL,Rigol2011}. Modern treatments discuss
the associated integral equations and exact thermodynamic
relations~\cite{Farina2022,Ristivojevic2026}. Thermodynamic
Bethe-ansatz calculations also benchmark analytical approximations
at finite temperature~\cite{Kerr2024}. The dimensionless coupling
is $\gamma_{\rm LL}=mg_0/(\hbar^2\rho)$, where $m$ is the atomic
mass, $g_0>0$ is the 1D coupling, and $\rho$ is the linear density.
At weak coupling, $\gamma_{\rm LL}\ll1$, the leading correction to
the mean-field energy is the negative Bogoliubov contribution,
commonly called the Lee--Huang--Yang (LHY)
term~\cite{Lieb1963,MoraCastin2003,Pitaevskii2016}. The LHY energy
is infrared convergent even though the condensate expansion
yields a logarithmically divergent depletion in the thermodynamic
limit~\cite{MoraCastin2003,Petrov2000}. The subsequent LL energy
coefficients provide a quantitative test of the terms generated
by a self-consistent field theory~\cite{Prolhac2017,Ristivojevic2019}.

The Cornwall--Jackiw--Tomboulis (CJT) effective action treats the
background field and the full propagator as independent
variables~\cite{Cornwall1974}. Stationarity gives equations for
the background and self-energy. At a finite
two-particle-irreducible (2PI) truncation, solving these equations
self-consistently resums classes of diagrams through the dressed
propagator~\cite{Baym1962,Cornwall1974}. The 2PI formulation
also provides a basis for constructing conserving approximations
and relating the self-energy to response
vertices~\cite{Baym1962,Griffin1996}. In three-dimensional (3D)
Bose gases, CJT calculations yield analytical EOS
expansions~\cite{Zhang2024}, and related Hartree--Fock (HF)
implementations address condensation and
depletion~\cite{Van2022}. A finite HF truncation can introduce a
spurious gap in the internal propagator, motivating a gapless
prescription~\cite{HugenholtzPines1959,Griffin1996}. Comparison
with the exact LL coefficients then tests the perturbative
accuracy of the resummation.

The infrared behavior of density and phase fluctuations poses a
second test in 1D and is central to the quasicondensate
description~\cite{Petrov2000,MoraCastin2003}. At zero temperature,
the homogeneous repulsive gas has algebraically decaying one-body
correlations and a vanishing condensate fraction in the
thermodynamic limit~\cite{Haldane1981PRL,Rigol2011}. Weak coupling
suppresses density fluctuations and permits an expansion about
a finite-size condensate or a locally coherent
background~\cite{Petrov2000,MoraCastin2003,Rigol2011}. In the
Cartesian-field representation, the transverse equal-time
correlator diverges logarithmically, while the longitudinal
correlator remains infrared integrable~\cite{MoraCastin2003}.
The depletion can therefore become appreciable at sufficiently
large length scales even at weak coupling. Extracting the bulk
EOS requires tracking this dependence on the infrared scale.

We apply the gapless double-bubble CJT approximation to a
single-component 1D Bose gas with repulsive contact interactions
at zero temperature. The background, Dyson, and number equations
eliminate the transverse correlator from the bulk density closure.
The longitudinal integral leads to a quadratic equation whose
positive root determines the dressed scale, chemical potential,
and energy density. We compare the resulting EOS with the LL
expansion and identify the first discrepancy at relative order
$\gamma_{\rm LL}$ beyond the common LHY term.

We evaluate the correlators with a common vacuum subtraction and
physical infrared cutoff. The depletion is nonnegative and depends
logarithmically on the size of the specified coherent region.
We distinguish the thermodynamic sound velocity from the
low-momentum slope of the internal CJT propagator and identify the
response calculation required to determine the physical collective
mode. Section~\ref{sec:cjt} presents the model and gapless
construction. Section~\ref{sec:eos} derives and benchmarks the EOS,
and Sec.~\ref{sec:ir} discusses the infrared interpretation of
depletion. Appendix~\ref{P11P22} details the fluctuation integrals.

\section{Model and gapless CJT construction}\label{sec:cjt}
\subsection{Action and fluctuation fields\label{cjt1}}
We consider bosons of mass $m$ on a line of length $L$ with
repulsive $\delta$-function interactions and linear density
$\rho=N/L$. The Euclidean path-integral
representation~\cite{Peskin1995,Nagaosa1999,Pitaevskii2016} gives
the grand-canonical partition function
\begin{equation}\label{partition function}
	\mathcal{Z}=\int\mathcal{D}\left[\mathbf{\Phi},\mathbf{\Phi}^{*}\right] \exp\left\{-\frac{S\left[\mathbf{\Phi},\mathbf{\Phi}^{*}\right]}{\hbar}\right\},
\end{equation}
with Euclidean action functional $S\left[\mathbf{\Phi},\mathbf{\Phi}^{*}\right]$ in Eq.~(\ref{partition function}) 
\begin{equation}\label{S}
\begin{split}
S[\mathbf{\Phi},\mathbf{\Phi}^{*}]
={}&\int_{0}^{\beta\hbar}\!d\tau\int_{0}^{L}\!dx\,
\biggl\{\mathbf{\Phi}^{*}(x,\tau)\\
&\times\biggl[\hbar\partial_{\tau}
-\frac{\hbar^{2}\partial_x^2}{2m}-\mu\biggr]
\mathbf{\Phi}(x,\tau)\\
&\qquad+\frac{g_0}{2}|\mathbf{\Phi}(x,\tau)|^4\biggr\}.
\end{split}
\end{equation}
In Equation~(\ref{S}), the complex field $\mathbf{\Phi}(x,\tau)$ is the bosonic field at position $x$
and imaginary time $\tau$, $\mu$ is the chemical potential, and
$\beta=(k_{\rm B}T)^{-1}$ is the inverse thermal energy, with
temperature $T$ and Boltzmann constant $k_{\rm B}$. The physical 1D contact coupling $g_0$ has units of energy times length~\cite{Sekino2021}. Long-wavelength fluctuations remain important at weak coupling and yield algebraic one-body correlations in the thermodynamic
limit~\cite{Haldane1981PRL,Petrov2000,Rigol2011}. For repulsive interactions, $g_0=-2\hbar^2/(ma_{\rm s})>0$ and
$a_{\rm s}<0$~\cite{Olshanii1998,Rigol2011}. The corresponding LL coupling parameter is
$\gamma_{\mathrm{LL}}=m g_{0}/(\hbar^{2}\rho)=2/\rho (|a_{\mathrm{s}}|)$.
Accordingly, weak coupling corresponds to
$\rho |a_{\mathrm{s}}|\gg 1$ and
$\gamma_{\mathrm{LL}}\ll 1$~\cite{Lieb1963,Rigol2011}.

The stationary effective potential associated with partition function~(\ref{partition function}) satisfies
$\mathcal{Z}=e^{-\beta L V_{\rm eff}}$. At one loop approximation, $V_{\rm eff}[\phi]$ depends solely on the background field $\phi$, which reprensents the expectation value of the quantum field $\hat{\mathbf{\Phi}}(x)$~\cite{MoraCastin2003,Pitaevskii2016}, and gives the LHY correction associated with
Eq.~\eqref{S}~\cite{MoraCastin2003,Pitaevskii2016}. 
To study higher-order EOS contributions generated by enhanced quantum fluctuations in 1D system, we employ CJT potential $V_{\rm eff}[\phi,G]$ to compute terms beyond LHY order. A double Legendre transform of the generating functional with linear and bilinear sources makes $\phi$ and the dressed connected propagator $G$ independent variables, where $G(u,v)$ corresponds to the expectation value of the time-ordered product $\mathcal{T}\hat{\mathbf{\Phi}}^{\dagger}(u)\hat{\mathbf{\Phi}}(v)$. Physical solutions are then identified by demanding that the generalized effective potential obey the stationarity conditions $\delta V_{\text{eff}}/\delta \phi=0$ and $\delta V_{\text{eff}}/\delta G=0$.
The CJT approach therefore rests on a self-consistent loop expansion of the effective potential $V_{\text{eff}}$, written in terms of the background field $\phi$ and the full propagator $G$~\cite{Cornwall1974,Baym1962}.

We then decompose $\mathbf{\Phi}$ in Eq.~\eqref{S} into a uniform real background
$\phi_0$ and two real fluctuation fields $\phi_1$ and $\phi_2$: $ \mathbf{\Phi}=(\phi_0+\phi_1+i\phi_2)/\sqrt{2}$.
The real background represents a finite-size condensate or a locally coherent quasicondensate. It does not imply spontaneous long-range order in the infinite 1D system~\cite{MoraCastin2003,Rigol2011}. A systematic derivation of $V_{\text{eff}}$ employs a double
Legendre transform of the action functional $S[\mathbf{\Phi},\mathbf{\Phi}^{*}]$ in Eq.~\eqref{S},
together with the stationarity conditions $\delta V_{\text{eff}}(\phi_{0},G) /\delta \phi_{0}=0$ and $\delta V_{\text{eff}}(\phi_{0},G)/\delta G=0$, in which $G$ is varied simultaneously with $\phi_0$. Through the effective potential $V_{\text{eff}}$, this construction incorporates both condensate and fluctuation contributions to the dynamics. 
Following the CJT formulation of Refs.~\cite{Cornwall1974,Song2022,Van2022,Zhang2024}, we Fourier transform the two fluctuation fields. With
$\epsilon_k=\hbar^2k^2/(2m)$ and Matsubara energy
$\omega_n=2\pi n/\beta$, the quadratic action gives the inverse Gaussian propagator
\begin{equation}
G_0^{-1}(k,\omega_n)=
\begin{pmatrix}
\epsilon_k-\mu+\frac{3g_0}{2}\phi_{0}^{2}&-\omega_n\\
\omega_n&\epsilon_k-\mu+\frac{g_0}{2}\phi_{0}^{2}
\end{pmatrix}.
\label{eq:G0-1}
\end{equation}
After analytical calculation, the pole of $G_0$ gives the Gaussian excitation spectrum
$\varepsilon_k=\sqrt{(\epsilon_k-\mu+3g_0\phi_0^2/2)
(\epsilon_k-\mu+g_0\phi_0^2/2)}$.

\subsection{Effective potential and restoration of a gapless mode\label{cjt2}}
In Section~\ref{cjt1}, we have introduced the action $S[\mathbf{\Phi},\mathbf{\Phi}^{*}]$ and fluctuation fields $\phi_{1}$, $\phi_{2}$ of the 1D system and led to the inverse Gaussian propagator $G_0^{-1}(k,\omega_n)$ at one-loop order. The CJT effective-field-theory framework is developed in Sec.~\ref{cjt2}. A double Legendre transform with respect to linear and bilinear sources produces the CJT potential $V_{\text{eff}}(\phi_{0},G)$. For the contact model, the double-bubble HF truncation takes the form
\begin{equation}
V_{\text{eff}}[\phi_0,G]=V_0+\frac12\int_\beta
\operatorname{Tr}\!\left[\ln G^{-1}+G_0^{-1}G-\mathbf1\right]
+\mathcal V_{2,\text{eff}},\label{eq:veff}
\end{equation}
with
$V_0=-\mu\rho_0+g_0\rho_0^2/2$, $\rho_0=\phi_0^2/2$,
$\mathcal V_{2,\text{eff}}=3g_0(P_{11}^2+P_{22}^2)/8
+g_0P_{11}P_{22}/4$.
In Equation~\eqref{eq:veff}, $G_0$ and $G$ are the Gaussian and dressed propagators,
respectively. The integrated diagonal components of $G$ are
\begin{equation}
	\label{p11&p22}
P_{aa}=\int_\beta G_{aa}, (a=1,2).\quad\int_\beta\equiv\frac1\beta\sum_{n=-\infty}^{\infty}
\int\frac{dk}{2\pi}.
\end{equation}
The trace runs over the two real field components, and the
logarithm contains an implicit field-independent reference.
Evaluating the double-bubble functional self-consistently sums a
subset of higher-order contributions.

Requiring the CJT effective potential $V_{\text{eff}}$ in Eq.~(\ref{eq:veff}) to be stationary ($\delta V_{\text{eff}}/\delta G=0$) with respect to the propagator components $G(k)$ yields $G^{-1}=G_0^{-1}+\Sigma$, where $\Sigma=2\delta \mathcal{V}_{2,\text{eff}}/\delta G$. In the unmodified approximation,
$\Sigma_1=3g_0P_{11}/2+g_0P_{22}/2$ and
$\Sigma_2=g_0P_{11}/2+3g_0P_{22}/2$.
The background equation is $\mu=g_0\rho_0+\Sigma_1$, obtained by imposing $\delta V_{\text{eff}}/\delta \phi_0=0$; consequently, the transverse inverse propagator at zero momentum equals $\Sigma_2-\Sigma_1$. 

To implement the gapless prescription of Ref.~\cite{Ivanov2005}, we introduce a backflow term 
$\Delta V=-g_0(P_{11}-P_{22})^2/4$.
The potential then becomes
\begin{eqnarray}
V_{\text{eff}}&=&V_0+\frac12\int_\beta
\operatorname{Tr}\!\left[\ln G^{-1}+G_0^{-1}G-\mathbf1\right]\nonumber\\
&\quad&+\frac{g_0}{8}(P_{11}^2+P_{22}^2)
+\frac{3g_0}{4}P_{11}P_{22}.
\label{eq:vg}
\end{eqnarray}
The backflow term implements the gapless prescription at fixed
microscopic coupling. It is independent of $\phi_0$ at fixed $G$,
so the background equation retains its form. The modified
diagonal self-energies are
$\Sigma_1^{\prime}=g_0P_{11}/2+3g_0P_{22}/2$ and
$\Sigma_2^{\prime}=3g_0P_{11}/2+g_0P_{22}/2$.
Since $\Sigma_2^{\prime}=\Sigma_1$, the background equation reads
\begin{equation}
	\label{gapeq}
\mu=g_0\rho_0+\Sigma^{\prime}_2.
\end{equation}
The Schwinger--Dyson (SD) equation then gives the longitudinal energy scale
\begin{equation}
	\label{M2}
M^2=-\mu+3g_0\rho_0+\Sigma^{\prime}_1.
\end{equation}
In Equation.~\eqref{M2}, $M^2$ carries units of energy and sets the dressed excitation scale. For the branch $M>0$, the full inverse propagator is
\begin{equation}
G^{-1}\left(k,n\right)=\begin{pmatrix}\epsilon_k+M^2&-\omega_n\\
\omega_n&\epsilon_k\end{pmatrix},\label{G}
\end{equation}
with excitation energy
$
\varepsilon_k=\sqrt{\epsilon_k(\epsilon_k+M^2)}.\label{eq:dispersion}
$
Accordingly, $\varepsilon_k\simeq\hbar |k|M/\sqrt{2m}$ at small momentum. The vanishing gap is a direct consequence of the background equation, as required by gaplessness~\cite{HugenholtzPines1959}.

At stationarity, the density follows from the explicit derivative
of the grand potential with respect to $\mu$. With vacuum
subtraction~\cite{Pitaevskii2016}, the number equation is
\begin{equation}
\rho=-\frac{d V_{\text{eff}}}{d\mu}
=\rho_0+\frac12(P_{11}+P_{22}).
\label{eq:number}
\end{equation}
Eliminating the background from Eqs.~(\ref{gapeq}), (\ref{M2}), and (\ref{eq:number}) yields
\begin{equation}
M^2=2g_0\rho-2g_0P_{11}.
\label{eq:closed}
\end{equation}
This relation is the central result furnished by the CJT effective potential $V_{\text{eff}}$. Although the phase integral $P_{22}$ is separately infrared sensitive in 1D, it drops out of these algebraic combinations. Interpreting the relations as a density EOS still requires that the background be assigned a finite-size or local-coherence meaning.

\section{1D integrals and EOS\label{sec:eos}}
Section~\ref{sec:cjt} combined the gap, SD, and number equations to derive the self-consistency equation for the parameter $M$ in Eq.~(\ref{eq:closed}). The present section (Sec.~\ref{sec:eos}) derives explicit analytical expressions for the correction beyond LHY order to the EOS of a 1D Bose gas. The starting point is the self-consistency equation for $M$, Eq.~(\ref{eq:closed}), obtained from the CJT effective potential $V_{\text{eff}}$, Eq.~(\ref{eq:vg}), in the two-loop approximation. The first task is to evaluate the zero-temperature fluctuation integrals ($P_{11}$ and $P_{22}$) for the 1D system. Inserting $P_{11}$ into Eq.~(\ref{eq:closed}) as a function of $M$ and solving the resulting closed equation then determines $M$ and yields the 1D beyond-LHY EOS.
\subsection{Zero-temperature fluctuation integrals\label{integral}}
We first express $P_{11}$ and $P_{22}$ in Eq.~(\ref{p11&p22}) as functions of $M$, carrying out both the sum over the Matsubara index $n$ and the integral over the wave vector $\mathbf{k}$.
The Matsubara sum and its zero-temperature limit yield
\begin{equation}
\frac{1}{\beta}\sum_n\frac{1}{\omega_n^2+\varepsilon_k^2}=\frac{1}{2\varepsilon_k}\coth\left(\frac{\beta \varepsilon_k}{2}\right)\xrightarrow[\beta\to\infty]{}\frac{1}{2\varepsilon_k},
\label{eq:matsubara_direct}
\end{equation}
with $\varepsilon_k=\sqrt{\epsilon_k(\epsilon_k+M^2)}$.
Accordingly, the two momentum integrals $P_{11}$ and $P_{22}$ that enter the number equation in Eq.~\eqref{eq:number} read
\begin{eqnarray}
P_{11}&=&\frac{1}{2}\int\frac{dk}{2\pi}\sqrt{\frac{\frac{\hbar^2k^2}{2m}}{\frac{\hbar^2k^2}{2m}+M^2}},\\
P_{22}&=&\frac{1}{2}\int\frac{dk}{2\pi}\sqrt{\frac{\frac{\hbar^2k^2}{2m}+M^2}{\frac{\hbar^2k^2}{2m}}}.
\label{eq:momentum_integrals_direct}
\end{eqnarray}
Integrating the vacuum-subtracted longitudinal kernel gives
the bulk 1D correlator [for the
detailed derivation, see Eqs.~\eqref{P11} and \eqref{P111D}
in Appendix~\ref{P11P22}],
\begin{equation}
	P_{11}^{(1\mathrm D)}=-\frac{q}{2\pi}
	=-\frac{\sqrt{m}\,M}{\sqrt{2}\,\pi\hbar},
	\qquad q\equiv\frac{\sqrt{2m}\,M}{\hbar}.
	\label{eq:p11_direct}
\end{equation}
The negative sign indicates that the longitudinal variance lies
below its free-particle vacuum reference. The particle occupation
involves the sum of the two subtracted correlators.

The transverse kernel scales as $q/|k|$ for $|k|\ll q$, giving
the infrared logarithm associated with phase
fluctuations~\cite{MoraCastin2003,Rigol2011}. We introduce a
physical lower cutoff $k_{\rm IR}>0$, of order $2\pi/L$ for a
finite periodic system or the inverse size of a locally coherent
region. Applying the same vacuum subtraction gives [for the detailed derivation, see Eqs.~\eqref{eq:P22} and \eqref{eq:app-p22-sub-exact} in Appendix~\ref{P11P22}]
\begin{equation}
	P_{22}^{\rm IR}(q;k_{\rm IR})
	=\frac{1}{2\pi}\int_{k_{\rm IR}}^\infty dk
	\left[\sqrt{1+\frac{q^2}{k^2}}-1\right].
	\label{eq:p22-ir-integral}
\end{equation}
Direct integration yields
\begin{equation}
		P_{22}^{\rm IR}
		=\frac{1}{2\pi}\left[
		k_{\rm IR}-\sqrt{k_{\rm IR}^2+q^2}
		+q\ln\!\left(\frac{q+\sqrt{k_{\rm IR}^2+q^2}}{k_{\rm IR}}\right)
		\right].
	\label{eq:p22-ir-exact}
\end{equation}
For $k_{\rm IR}\ll q$,
\begin{equation}
	P_{22}^{\rm IR}
	=\frac{q}{2\pi}\left[\ln\!\left(\frac{2q}{k_{\rm IR}}\right)-1\right]
	+\frac{k_{\rm IR}}{2\pi}
	+O\!\left(\frac{k_{\rm IR}^2}{q}\right).
	\label{eq:p22-ir-asymptotic}
\end{equation}
The regime $k_{\rm IR}\ll q$ makes the argument of the logarithm
large. Keeping the logarithm preserves the infrared asymptotics.
The physical cutoff also defines the coherent region used to
evaluate the depletion.

\subsection{Self-consistent determination of $M$\label{Meq}}
Following the evaluation of the Matsubara sum and the two momentum integrals at zero temperature in Sec.~\ref{integral}, we turn to Sec.~\ref{Meq} and obtain analytical expressions for the beyond-LHY EOS of the 1D Bose gas in terms of $M$. Substituting Eq.~(\ref{eq:p11_direct}) into the closed equation yields
\begin{equation}
M^2-\frac{\sqrt{2m}\,g_0}{\pi\hbar}M-2g_0\rho=0.
\label{eq:quadratic_direct}
\end{equation}
The physically admissible branch is the positive root,
\begin{equation}
	M=\frac{A+\sqrt{A^2+8g_0\rho}}{2},
	\qquad A\equiv\frac{\sqrt{2m}\,g_0}{\pi\hbar}.
	\label{eq:M-exact}
\end{equation}
Within the repulsive convention
$g_0=-2\hbar^2/(ma_s)>0$ with $a_s<0$,
$\gamma_{\rm LL}=mg_0/(\hbar^2\rho)=2/(\rho|a_s|)\ll1
$,
which implies
\begin{equation}
	M=\sqrt{2g_0\rho}\left[1+
	\frac{\sqrt{\gamma_{\rm LL}}}{2\pi}
	+\frac{\gamma_{\rm LL}}{8\pi^2}
	+\mathcal{O}(\gamma_{\rm LL}^{2})\right].
	\label{eq:MEQ}
\end{equation}
The positive signs are a direct consequence of $P_{11}<0$ in Eq.~\eqref{eq:p11_direct}.
Substituting the leading term $M=\sqrt{2g_0\rho}$ into the fluctuation correction to the chemical potential $\mu$ generates the LHY contribution.
Keeping the successive powers of $\sqrt{\gamma_{\rm LL}}$ produces the higher-order terms of the chosen CJT resummation. For $\gamma_{\rm LL}\ll1$, the hierarchy
$\gamma_{\rm LL}\ll\sqrt{\gamma_{\rm LL}}\ll1$ holds, and its quantitative accuracy is examined against the exact energy below.

With $M$ determined by Eq.~(\ref{eq:MEQ}), the beyond-LHY EOS of the model can be evaluated from the CJT effective potential in Eq.~(\ref{eq:vg}). We focus on three quantities: the quantum depletion $\rho_{\text{ex}}$, the chemical potential $\mu$, and the energy density $\mathcal{E}$.

\subsection{Fluctuations and quantum depletion}
To begin, with the aim of calculating the quantum depletion $\rho_{\text{ex}}$, the fluctuations in Eqs.~(\ref{eq:p11_direct}) and~(\ref{eq:p22-ir-asymptotic}) admit expansions in the LL coupling parameter $\gamma_{\text{LL}}$. Inserting Eq.~(\ref{eq:MEQ}) into Eqs.~(\ref{eq:p11_direct}) and~(\ref{eq:p22-ir-asymptotic}), respectively, yields
\begin{equation}
	P_{11}^{(1\mathrm D)}=-\rho\left[
	\frac{\gamma_{\rm LL}^{1/2}}{\pi}
	+\frac{\gamma_{\rm LL}}{2\pi^2}
	+\frac{\gamma_{\rm LL}^{3/2}}{8\pi^3}
	+\mathcal{O}(\gamma_{\rm LL}^{2})\right].
	\label{eq:P11LL}
\end{equation}
To expose the cutoff dependence, we define
\begin{equation}
	\Lambda_{\rm IR}\equiv
	\ln\!\left(\frac{4\rho\sqrt{\gamma_{\rm LL}}}{k_{\rm IR}}\right),
	\qquad k_{\rm IR}\ll2\rho\sqrt{\gamma_{\rm LL}}.
	\label{eq:LambdaIR}
\end{equation}
A consistent expansion of Eq.~\eqref{eq:p22-ir-asymptotic} in
$\gamma_{\rm LL}$ yields
\begin{eqnarray}
	\frac{P_{22}^{\rm IR}}{\rho}
	&=&\frac{\gamma_{\rm LL}^{1/2}}{\pi}(\Lambda_{\rm IR}-1)
	+\frac{\gamma_{\rm LL}}{2\pi^2}\Lambda_{\rm IR}
	\nonumber\\
	&+&\frac{\gamma_{\rm LL}^{3/2}}{8\pi^3}(\Lambda_{\rm IR}+1)
	+\mathcal{O}(\gamma_{\rm LL}^{2})
	+\mathcal{O}\!\left(\frac{k_{\rm IR}}{\rho}\right).
	\label{eq:P22LL}
\end{eqnarray}
The depletion is the fluctuation contribution to the number equation. For a common physical cutoff, we use the longitudinal result of Eq.~\eqref{P11} together with the transverse result of Eq.~\eqref{eq:p22-ir-exact},
\begin{equation}
    \rho_{\rm ex}^{\rm IR}\equiv
    \frac{P_{11}^{\rm IR}+P_{22}^{\rm IR}}{2}.
    \label{eq:rhoex-definition}
\end{equation}
For a sharp continuum cutoff this gives
\begin{equation}
\begin{split}
\rho_{\rm ex}^{\rm IR}
={}&\frac{1}{4\pi}\biggl[2k_{\rm IR}-2\sqrt{k_{\rm IR}^2+q^2}\\
&\qquad+q\ln\!\left(\frac{q+\sqrt{k_{\rm IR}^2+q^2}}{k_{\rm IR}}\right)\biggr].
\end{split}
\label{eq:rhoex-exact}
\end{equation}
For $k_{\rm IR}\ll q$, the difference between $P_{11}^{\rm IR}$
and $P_{11}^{(1\mathrm D)}$ is $\mathcal{O}(k_{\rm IR})$. With this
correction included in the cutoff remainder,
Eqs.~\eqref{eq:P11LL} and \eqref{eq:P22LL} give
\begin{eqnarray}
	\frac{\rho_{\rm ex}^{\rm IR}}{\rho}
	&=&\frac{\gamma_{\rm LL}^{1/2}}{2\pi}(\Lambda_{\rm IR}-2)
	+\frac{\gamma_{\rm LL}}{4\pi^2}(\Lambda_{\rm IR}-1)
	\nonumber\\
	&+&\frac{\gamma_{\rm LL}^{3/2}}{16\pi^3}\Lambda_{\rm IR}
	+\mathcal{O}(\gamma_{\rm LL}^{2})
	+\mathcal{O}\!\left(\frac{k_{\rm IR}}{\rho}\right).
	\label{eq:rhoex-gamma}
\end{eqnarray}
Both $\gamma_{\rm LL}\ll1$ and
$\rho_{\rm ex}^{\rm IR}/\rho\ll1$ are necessary for a controlled fluctuation expansion. At fixed density, $\rho_{\rm ex}^{\rm IR}$ grows logarithmically when $k_{\rm IR}\to0$. The infinite homogeneous 1D gas therefore possesses no finite global condensate depletion; the CJT background should instead be understood as a local coherent (quasicondensate) background.

\subsection{Chemical potential and energy density}
Combining the background equation, Eq.~\eqref{gapeq}, with the
number equation, Eq.~\eqref{eq:number}, eliminates $P_{22}$ from
the chemical potential and yields
\begin{equation}
\label{muM}
\mu=g_{0}\rho-g_{0}\frac{m^{1/2}M}{\sqrt{2}\pi\hbar}.
\end{equation}
Expanding $M$ according to Eq.~\eqref{eq:MEQ} gives
\begin{equation}
	\label{mugamma}
	\mu=g_0\rho\left[
	1-\frac{\gamma_{\text{LL}}^{1/2}}{\pi}
	-\frac{\gamma_{\text{LL}}}{2\pi^2}
	-\frac{\gamma_{\text{LL}}^{3/2}}{8\pi^3}
	+\mathcal{O}\left(\gamma_{\text{LL}}^{2}\right)
	\right].
\end{equation}
At fixed $g_0$, the energy density follows from
$d\mathcal E/d\rho=\mu$. With the density-independent integration
constant set to zero, the weak-coupling expansion is
\begin{align}
	\mathcal E
	=\frac{g_0\rho^2}{2}\biggl[1
	&-\frac{4}{3\pi}\sqrt{\gamma_{\rm LL}}
	-\frac{\gamma_{\rm LL}}{\pi^2}
	-\frac{\gamma_{\rm LL}^{3/2}}{2\pi^3}
	+\mathcal{O}(\gamma_{\rm LL}^{2})\biggr].
	\label{eq:energy-gamma}
\end{align}
Equation~\eqref{eq:energy-gamma} admits an order-by-order benchmark
against the ground-state energy of the repulsive Lieb--Liniger gas.
Within the same coupling convention, the weak-coupling expansion of the Bethe-ansatz solution~\cite{Lieb1963,Prolhac2017,Ristivojevic2019} takes the form
\begin{equation}
	\begin{split}
		\mathcal E_{\rm LL}=\frac{g_0\rho^2}{2}\biggl[1
		&-\frac{4}{3\pi}\sqrt{\gamma_{\rm LL}}\\
		&+\left(\frac16-\frac{1}{\pi^2}\right)\gamma_{\rm LL}
		+\mathcal{O}(\gamma_{\rm LL}^{3/2})\biggr].
	\end{split}
	\label{eq:energy-LL-benchmark}
\end{equation}
Both expansions share the mean-field limit $g_0\rho^2/2$ and
the relative LHY correction
$-4\sqrt{\gamma_{\rm LL}}/(3\pi)$~\cite{MoraCastin2003,Pitaevskii2016}.
The zero-point energy correction remains infrared convergent in
the thermodynamic limit. The first discrepancy occurs at relative
order $\gamma_{\rm LL}$. The CJT coefficient is $-1/\pi^2$, and
the exact coefficient is $1/6-1/\pi^2$. Their difference gives
\begin{equation}
	\mathcal E-\mathcal E_{\rm LL}
	=-\frac{g_0\rho^2}{12}\gamma_{\rm LL}
	+\mathcal{O}\!\left(g_0\rho^2\gamma_{\rm LL}^{3/2}\right).
	\label{eq:energy-benchmark-difference}
\end{equation}
In units of $\hbar^2\rho^3/(2m)$, the CJT energy thus agrees
with the exact LL expansion through $\gamma_{\rm LL}^{3/2}$
and first differs at $\gamma_{\rm LL}^2$. The subsequent
coefficients in Eq.~\eqref{eq:energy-gamma} are predictions of
the selected double-bubble resummation.

\begin{figure}[t]
  \centering
  \includegraphics[width=\columnwidth]{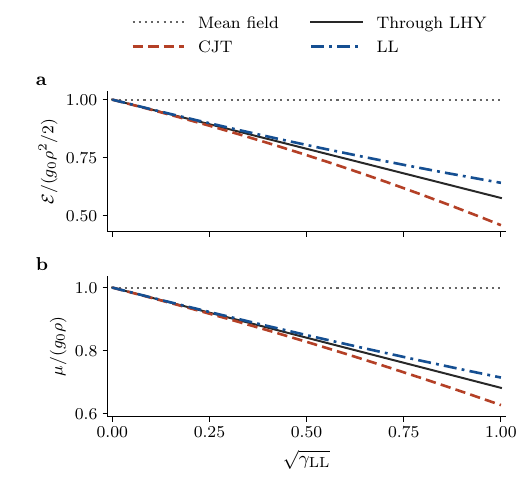}
  \caption{Zero-temperature EOS as a function of
  $\sqrt{\gamma_{\rm LL}}$.
  (a) Energy density $\mathcal E$ in units of $g_0\rho^2/2$.
  (b) Chemical potential $\mu$ in units of $g_0\rho$.
  The curves show mean field (gray dotted), the expansion through
  LHY order (black solid), the CJT expansion through relative order
  $\gamma_{\rm LL}^{3/2}$ (red dashed), and the LL expansion through
  relative order $\gamma_{\rm LL}$ (blue dash-dotted).
  The chemical potentials follow from the corresponding energy
  densities at fixed $g_0$.}
  \label{fig:eos-benchmark}
\end{figure}

Figure~\ref{fig:eos-benchmark}(a) shows that the common LHY
correction lowers the energy density below $g_0\rho^2/2$.
At the next order, the CJT curve falls below the LHY baseline
and the LL curve lies above it. For the chemical potential,
each relative energy term proportional to $\gamma_{\rm LL}^p$
acquires a factor $1-p/2$ under differentiation at fixed $g_0$,
because $\gamma_{\rm LL}\propto\rho^{-1}$. Both chemical-potential
curves therefore share the LHY term
$-\sqrt{\gamma_{\rm LL}}/\pi$. Their leading difference in
Fig.~\ref{fig:eos-benchmark}(b) is
$(\mu-\mu_{\rm LL})/(g_0\rho)=-\gamma_{\rm LL}/12
+O(\gamma_{\rm LL}^{3/2})$. Near $\gamma_{\rm LL}=1$,
the curves illustrate the displayed truncations outside
the asymptotic weak-coupling regime.

\subsection{Density response: sound velocity $c_{\rm th}$ and $c_{\rm sp}$}
Equation~\eqref{eq:energy-gamma} embodies the density-dependent
energy through the displayed order. 
The first correction is the negative 1D Bogoliubov (LHY)
zero-point contribution; the later terms arise from the selected gapless HF resummation. Their first departure from the exact LL expansion is quantified in Eq.~\eqref{eq:energy-benchmark-difference}.

The EOS also determines the thermodynamic sound velocity
$c_{\rm th}^2=(\rho/m)\,\partial_\rho\mu$. At fixed $g_0$,
Eq.~\eqref{mugamma} gives
\begin{equation}
	c_{\rm th}^2=\frac{g_0\rho}{m}\left[1-
	\frac{\sqrt{\gamma_{\rm LL}}}{2\pi}
	+\frac{\gamma_{\rm LL}^{3/2}}{16\pi^3}
	+O(\gamma_{\rm LL}^{2})\right].
	\label{eq:cth}
\end{equation}
The term proportional to $\gamma_{\rm LL}$ in the normalized
chemical potential gives a density-independent contribution to
$\mu$ at fixed $g_0$ and drops out of $\partial_\rho\mu$.
At zero temperature, the inverse isothermal compressibility is
$\rho^2\partial_\rho\mu=m\rho c_{\rm th}^2$.
The LL EOS can also be used in hydrodynamic calculations of
long-wavelength excitations~\cite{Kopycinski2022}. Within the
CJT calculation, the internal propagator gives
$c_{\rm sp}=\lim_{k\to0}\varepsilon_k/(\hbar|k|)
=M/\sqrt{2m}$, with weak-coupling expansion
\begin{equation}
	c_{\rm sp}^{2}=\frac{g_0\rho}{m}\left[1+
	\frac{\sqrt{\gamma_{\rm LL}}}{\pi}
	+\frac{\gamma_{\rm LL}}{2\pi^{2}}
	+\frac{\gamma_{\rm LL}^{3/2}}{8\pi^3}
	+O(\gamma_{\rm LL}^{2})\right].
	\label{eq:csp}
\end{equation}
The normalized squares of $c_{\rm th}$ and $c_{\rm sp}$ already
differ at order $\sqrt{\gamma_{\rm LL}}$. Stationarity and
gaplessness constrain the internal self-energy. Determining
the physical collective mode requires a density response with
vertices consistent with the effective-action truncation and
the corresponding Ward identities~\cite{Baym1962,Griffin1996}.

\section{Infrared fluctuations and the meaning of depletion}\label{sec:ir}
Section~\ref{sec:eos} derived a finite bulk EOS by eliminating
the transverse correlator. We now examine how the same correlator
controls the dependence of the depletion on the infrared scale.

For $|k|\ll q$, the longitudinal kernel in
Eq.~\eqref{eq:momentum_integrals_direct} vanishes as $|k|/q$,
while the transverse kernel diverges as $q/|k|$ and gives the
logarithm in Eq.~\eqref{eq:p22-ir-asymptotic}. At linear order
the phase and density fluctuations are
$\theta\simeq\phi_2/\phi_0$ and $\delta\rho\simeq\phi_0\phi_1$,
respectively.

We adopt a symmetric continuum cutoff $|k|\geq k_{\rm IR}$ that excludes the background mode and subtract the momentum-independent vacuum term in both channels. In particular,
\begin{align}
	P_{11}^{\rm IR}(k_{\rm IR})
	&=\frac{1}{2\pi}\int_{k_{\rm IR}}^\infty dk
	\left[\frac{k}{\sqrt{k^2+q^2}}-1\right]\nonumber\\
	&=\frac{k_{\rm IR}-\sqrt{k_{\rm IR}^2+q^2}}{2\pi}.
	\label{eq:p11cut}
\end{align}
Combining this result with Eq.~\eqref{eq:p22-ir-exact} yields
Eq.~\eqref{eq:rhoex-exact}. Equivalently, the depletion is the
integral of the nonnegative Bogoliubov occupation kernel,
\begin{equation}
	\begin{split}
		\rho_{\rm ex}^{\rm IR}
		=\int_{|k|\geq k_{\rm IR}}\frac{dk}{2\pi}\,
		\frac14\biggl[&
		\sqrt{\frac{\epsilon_k}{\epsilon_k+M^2}}\\
		&+\sqrt{\frac{\epsilon_k+M^2}{\epsilon_k}}-2\biggr].
	\end{split}
	\label{eq:depletion-occupation}
\end{equation}
Hence the negative vacuum-subtracted $P_{11}$ is consistent with a positive depletion. For $k_{\rm IR}\ll q$,
\begin{equation}
	\rho_{\rm ex}^{\rm IR}
	=\frac{q}{4\pi}\left[\ln\frac{2q}{k_{\rm IR}}-2\right]
	+\frac{k_{\rm IR}}{2\pi}
	+\mathcal{O}\!\left(\frac{k_{\rm IR}^2}{q}\right).
	\label{eq:depletionlog}
\end{equation}
The constant $-2$ originates from the specified vacuum subtraction and sharp continuum cutoff. Either a discrete finite-size sum or a different infrared window alters the finite constant while preserving the logarithmic coefficient at fixed $q$.

In a periodic system of length $L$, the lowest nonzero momentum is
$2\pi/L$. The continuum cutoff reproduces the logarithmic
coefficient, and matching to the discrete sum fixes the finite
constant. For a specified coherent region of size $\ell_{\rm coh}$,
one may choose $k_{\rm IR}\sim\ell_{\rm coh}^{-1}$. This length
characterizes the selected region. In the homogeneous
zero-temperature gas, the one-body correlations decay
algebraically~\cite{Haldane1981PRL,Rigol2011}.

The bulk solution in Eq.~\eqref{eq:M-exact} relies on
$P_{11}^{\rm IR}(0)=-q/(2\pi)$. If a finite cutoff is retained,
\begin{equation}
	P_{11}^{\rm IR}(k_{\rm IR})
	=-\frac{q}{2\pi}+\frac{k_{\rm IR}}{2\pi}
	+\mathcal{O}(k_{\rm IR}^2/q),
\end{equation}
so the bulk closure omits a relative correction of order
$k_{\rm IR}/q$ in the longitudinal fluctuation contribution. A fully finite-size self-consistent treatment would instead substitute Eq.~\eqref{eq:p11cut} into Eq.~\eqref{eq:closed}. Using the bulk $M$ in the depletion thus keeps the infrared logarithm while neglecting these finite-size corrections to the closure.

Equation~\eqref{eq:rhoex-gamma} expands both the prefactor $q$
and the argument of $\ln(2q/k_{\rm IR})$ about the same bulk solution for $M$. It applies for $\gamma_{\rm LL}\ll1$, $k_{\rm IR}\ll q$, and $\rho_{\rm ex}^{\rm IR}/\rho\ll1$. Its subleading coefficients depend on the selected HF resummation.

At fixed coupling and density, the logarithmic growth as
$k_{\rm IR}\to0$ signals the breakdown of the globally coherent expansion. The depletion is defined relative to a finite-size condensate or a specified local background. In the thermodynamic limit, the condensate fraction vanishes and correlation functions describe long-distance coherence. A quasicondensate density defined through suppressed density fluctuations is a distinct quantity~\cite{Petrov2000,MoraCastin2003,Rigol2011}.
Recent studies of finite-temperature correlations in the homogeneous gas~\cite{DeRosi2023} and of one-body correlations and momentum distributions in trapped gases~\cite{Takacs2025} further illustrate
why coherence should be characterized through correlation functions. Those settings include thermal or trapping effects beyond the homogeneous zero-temperature calculation considered here.

The coherent region fixes the physical infrared cutoff and hence the infrared dependence of the depletion. The algebraic cancellation of $P_{22}$ in Eqs.~\eqref{eq:closed} and \eqref{muM} keeps the bulk EOS finite as the cutoff vanishes. At finite size, corrections to $P_{11}^{\rm IR}$ can still shift the EOS if retained in the self-consistency equation.

\section{Conclusion and outlook}\label{sec:conclusion}
Within a specified gapless Hartree--Fock CJT approximation, we have derived the zero-temperature EOS of a repulsive 1D contact Bose gas. Removal of the background density from the stationarity and
number equations eliminates the infrared-singular transverse correlator from the bulk closure. The resulting quadratic equation admits a closed-form positive-root solution, which determines the chemical
potential $\mu$ and the energy density $\mathcal{E}$. Their weak-coupling expansions reproduce mean-field theory and the negative 1D LHY energy correction.

The LL benchmark also exposes the first quantitative
limitation of this construction. At relative order
$\gamma_{\rm LL}$ in the energy, the CJT coefficient is $-1/\pi^2$, while the exact value is $1/6-1/\pi^2$. The approximation therefore agrees through LHY order but fails to reproduce the complete next correction. Its remaining coefficients describe the selected self-consistent resummation and should be judged accordingly. Neither gaplessness nor self-consistency alone guarantees the exact EOS beyond the leading fluctuation contribution.

The infrared status of the depletion differs from that of the bulk energy. Applying the same vacuum subtraction to both correlators eliminates their ultraviolet parts, whereas the long-wavelength phase fluctuations
survive because of the physical lower momentum scale. The resulting nonnegative depletion is meaningful relative to a finite-size or specified local background
as long as its fraction remains small. Its logarithmic growth as the infrared cutoff vanishes signals the breakdown of a global condensate expansion, consistent with the absence of thermodynamic condensate order in 1D.

A controlled improvement beyond LHY order calls for additional contributions that are necessary to reproduce the exact weak-coupling expansion. Similarly, identification of the physical collective mode
requires a density response with vertices consistent with the effective-action truncation, rather than the assumption that it equals the internal propagator. These benchmarks and infrared prescriptions offer concrete tests for extending the CJT treatment of 1D Bose gases.

\section{Acknowledgments}
We are grateful to Tao Yu, Ying Hu, and Biao Wu for stimulating discussions and helpful assistance. This work was supported by the
National Natural Science Foundation of China under Grant No. 12574301 and the Zhejiang Provincial Natural Science Foundation of China under Grant No. LZ25A040004.

\appendix
\section{Detailed evaluation of the 1D fluctuation integrals}
\label{P11P22}
We evaluate the 1D equal-time correlators at zero temperature
with $M>0$ and $q=\sqrt{2m}M/\hbar$. Both integrals initially
use $k_{\rm IR}\leq|k|\leq k_{\rm UV}$, with
$0<k_{\rm IR}<k_{\rm UV}$. We remove the upper cutoff after
subtracting the shared vacuum contribution. The lower cutoff
specifies the physical infrared scale. The labels
\lq\lq reg\rq\rq\ and
\lq\lq IR\rq\rq\ denote the unsubtracted
and vacuum-subtracted expressions, respectively, at fixed
$k_{\rm IR}$.

\subsection{Longitudinal correlator: frequency integration and convergence}
For the longitudinal channel, the inverse of Eq.~\eqref{G} yields
\begin{equation}
	G_{11}(k,\omega)
	=\frac{\epsilon_k}{\omega^2+\varepsilon_k^2},
	\qquad
	\varepsilon_k^2=\epsilon_k(\epsilon_k+M^2).
	\label{eq:app-g11}
\end{equation}
Here $\omega$ carries units of energy, consistent with the Matsubara convention adopted in the main text. With
\begin{equation}
	\int_{-\infty}^{\infty}\frac{d\omega}{2\pi}
	\frac{1}{\omega^2+\varepsilon_k^2}
	=\frac{1}{2\varepsilon_k},
	\qquad k\neq0,
	\label{eq:app-frequency-integral}
\end{equation}
the zero-temperature equal-time correlator takes the form
\begin{eqnarray}
	P_{11}^{(1\mathrm D)}
	&=&\int_{-\infty}^{\infty}\frac{dk}{2\pi}
	\int_{-\infty}^{\infty}\frac{d\omega}{2\pi}
	\frac{\epsilon_k}{\omega^2+\varepsilon_k^2}\nonumber\\
	&=&\frac12\int_{-\infty}^{\infty}\frac{dk}{2\pi}
	\frac{\epsilon_k}{\varepsilon_k}\nonumber\\
	&=&\frac12\int_{-\infty}^{\infty}\frac{dk}{2\pi}
	\sqrt{\frac{\epsilon_k}{\epsilon_k+M^2}}\nonumber\\
	&=&\frac{1}{2\pi}\int_0^\infty dk\,
	\frac{k}{\sqrt{k^2+q^2}}.
	\label{eq:app-p11-raw}
\end{eqnarray}
As in the transverse integral below, the final line combines the two momentum directions. At large momentum,
\begin{equation}
	\frac{k}{\sqrt{k^2+q^2}}
	=1-\frac{q^2}{2k^2}+\mathcal{O}(q^4/k^4),
	\qquad k\gg q.
	\label{eq:app-p11-uv}
\end{equation}
The constant term generates a linear ultraviolet divergence. This term
is the free-particle vacuum contribution of $1/2$ per momentum mode, whereas the interaction-dependent remainder is ultraviolet integrable in 1D. For small positive momentum, by contrast,
\begin{equation}
	\frac{k}{\sqrt{k^2+q^2}}
	=\frac{k}{q}-\frac{k^3}{2q^3}+\mathcal{O}(k^5/q^5).
	\label{eq:app-p11-ir}
\end{equation}
At small momentum, the frequency integral scales as
$1/\varepsilon_k\propto1/|k|$. The longitudinal numerator
$\epsilon_k\propto k^2$ yields an integrable kernel proportional
to $|k|$. In the transverse channel, the numerator approaches
$M^2$ and the $1/|k|$ singularity survives.

\subsection{Longitudinal momentum integral and vacuum subtraction}
Under the common two-cutoff prescription,
\begin{equation}
	P_{11}^{\rm reg}
	=\frac{1}{2\pi}\int_{k_{\rm IR}}^{k_{\rm UV}}
	dk\,\frac{k}{\sqrt{k^2+q^2}}.
\end{equation}
The substitution $s=k^2+q^2$, $ds=2k\,dk$, gives
\begin{eqnarray}
	\int dk\,\frac{k}{\sqrt{k^2+q^2}}
	&=&\frac12\int ds\,s^{-1/2}\nonumber\\
	&=&\sqrt{s}=\sqrt{k^2+q^2},
\end{eqnarray}
apart from an integration constant. Consequently,
\begin{equation}
	P_{11}^{\rm reg}
	=\frac{\sqrt{k_{\rm UV}^2+q^2}
		-\sqrt{k_{\rm IR}^2+q^2}}{2\pi}.
	\label{eq:app-p11-reg-exact}
\end{equation}
For $k_{\rm UV}\gg q\gg k_{\rm IR}$, the endpoint expansion reads
\begin{equation}
	\begin{split}
		P_{11}^{\rm reg}
		={}&\frac{k_{\rm UV}}{2\pi}-\frac{q}{2\pi}
		+\frac{q^2}{4\pi k_{\rm UV}}
		-\frac{k_{\rm IR}^2}{4\pi q}\\
		&+\mathcal{O}\!\left(\frac{q^4}{k_{\rm UV}^3},
		\frac{k_{\rm IR}^4}{q^3}\right).
	\end{split}
\end{equation}
Only the first term is divergent. Subtraction of the free-particle vacuum variance with the same momentum limits yields
\begin{eqnarray}
	P_{11}^{\rm IR}
	&\equiv&\lim_{k_{\rm UV}\to\infty}
	\left[P_{11}^{\rm reg}
	-\frac{k_{\rm UV}-k_{\rm IR}}{2\pi}\right]\nonumber\\
	&=&\frac{1}{2\pi}\int_{k_{\rm IR}}^\infty dk
	\left(\frac{k}{\sqrt{k^2+q^2}}-1\right)\nonumber\\
	&=&\frac{1}{2\pi}\left[\sqrt{k^2+q^2}-k
	\right]_{k_{\rm IR}}^\infty\nonumber\\
	&=&\frac{k_{\rm IR}-\sqrt{k_{\rm IR}^2+q^2}}{2\pi}.
	\label{P11}
\end{eqnarray}
The upper endpoint vanishes because
$\sqrt{k^2+q^2}-k=q^2/(\sqrt{k^2+q^2}+k)\to0$.
Since the subtracted kernel tends to $-1$ at the lower endpoint, it remains infrared integrable. For $k_{\rm IR}\ll q$,
\begin{equation}
	P_{11}^{\rm IR}
	=-\frac{q}{2\pi}+\frac{k_{\rm IR}}{2\pi}
	-\frac{k_{\rm IR}^2}{4\pi q}
	+\mathcal{O}(k_{\rm IR}^4/q^3).
	\label{eq:app-p11-small-cutoff}
\end{equation}
The finite bulk limit therefore reads
\begin{equation}
	P_{11}^{(1\mathrm D),\rm sub}
	\equiv P_{11}^{\rm IR}(0)
	=-\frac{q}{2\pi}
	=-\frac{\sqrt m\,M}{\sqrt2\pi\hbar}.
	\label{P111D}
\end{equation}
Equation~\eqref{eq:p11_direct} uses this vacuum-subtracted bulk
value. Its negative sign reflects the reduction of the
longitudinal variance relative to the free-particle vacuum.
The finite-cutoff depletion uses the same nonzero $k_{\rm IR}$
in both channels. Expanding the bulk result using
Eq.~\eqref{eq:MEQ} gives Eq.~\eqref{eq:P11LL}.

\subsection{Equal-time transverse correlator and its endpoint behavior}

Inversion of Eq.~\eqref{G} gives
\begin{equation}
	G_{22}(k,\omega)
	=\frac{\epsilon_k+M^2}{\omega^2+\varepsilon_k^2},
	\qquad
	\varepsilon_k^2=\epsilon_k(\epsilon_k+M^2).
\end{equation}
The zero-temperature frequency integration then produces
\begin{eqnarray}
	P_{22}^{(1\mathrm D)}
	&=&\int_{-\infty}^{\infty}\frac{dk}{2\pi}
	\int_{-\infty}^{\infty}\frac{d\omega}{2\pi}
	\frac{\epsilon_k+M^2}{\omega^2+\varepsilon_k^2}
	\nonumber\\
	&=&\frac12\int_{-\infty}^{\infty}\frac{dk}{2\pi}
	\frac{\epsilon_k+M^2}{\varepsilon_k}
	\nonumber\\
	&=&\frac12\int_{-\infty}^{\infty}\frac{dk}{2\pi}
	\sqrt{\frac{\epsilon_k+M^2}{\epsilon_k}}
	\nonumber\\
	&=&\frac{1}{2\pi}\int_0^\infty dk\,
	\frac{\sqrt{k^2+q^2}}{k}.
	\label{eq:app-p22-raw}
\end{eqnarray}
The last equality exploits the evenness of the integrand. Both endpoints deserve attention. In the ultraviolet,
\begin{equation}
	\frac{\sqrt{k^2+q^2}}{k}
	=1+\frac{q^2}{2k^2}+\mathcal{O}(q^4/k^4),
	\qquad k\gg q.
	\label{eq:app-p22-uv}
\end{equation}
The constant term produces the linear ultraviolet divergence
and corresponds to the free-particle vacuum variance of $1/2$
per momentum mode in each real field component. The remaining
terms are ultraviolet integrable in 1D.

In the infrared, by contrast,
\begin{equation}
	\frac{\sqrt{k^2+q^2}}{k}
	=\frac{q}{k}+\frac{k}{2q}+\mathcal{O}(k^3/q^3),
	\qquad 0<k\ll q.
	\label{eq:app-p22-ir}
\end{equation}
The resulting $\int dk/k$ divergence originates from long-wavelength phase fluctuations of the gapless mode. It depends on the interaction and persists after subtraction of the free vacuum contribution.

\subsection{Integration with explicit ultraviolet and infrared cutoffs}

Applying the same two-cutoff prescription used for the longitudinal channel casts Eq.~\eqref{eq:app-p22-raw} as
\begin{equation}
	P_{22}^{\rm reg}
	=\frac{1}{2\pi}\int_{k_{\rm IR}}^{k_{\rm UV}}
	dk\,\frac{\sqrt{k^2+q^2}}{k}.
	\label{eq:app-p22-two-cutoffs}
\end{equation}
To evaluate it explicitly, set $k=q\sinh t$, so that
$dk=q\cosh t\,dt$ and $\sqrt{k^2+q^2}=q\cosh t$. This gives
\begin{eqnarray}
	\int dk\,\frac{\sqrt{k^2+q^2}}{k}
	&=&q\int dt\,\frac{\cosh^2t}{\sinh t}\nonumber\\
	&=&q\int dt\left(\sinh t+\frac{1}{\sinh t}\right)\nonumber\\
	&=&q\cosh t+q\ln\tanh\frac{t}{2}\nonumber\\
	&=&\sqrt{k^2+q^2}
	-q\ln\frac{q+\sqrt{k^2+q^2}}{k}.\nonumber\\
	\label{eq:app-p22-primitive}
\end{eqnarray}
Using $\tanh(t/2)=k/(q+\sqrt{k^2+q^2})$, we omit the integration
constant. Evaluation at the two endpoints yields
\begin{equation}
	\begin{split}
		2\pi P_{22}^{\rm reg}
		={}&\sqrt{k_{\rm UV}^2+q^2}-\sqrt{k_{\rm IR}^2+q^2}\\
		&-q\ln\frac{q+\sqrt{k_{\rm UV}^2+q^2}}{k_{\rm UV}}\\
		&+q\ln\frac{q+\sqrt{k_{\rm IR}^2+q^2}}{k_{\rm IR}}.
	\end{split}
	\label{eq:app-p22-reg-exact}
\end{equation}
For $k_{\rm UV}\gg q\gg k_{\rm IR}$, this expression becomes
\begin{equation}
	\begin{split}
		P_{22}^{\rm reg}
		={}&\frac{k_{\rm UV}}{2\pi}
		+\frac{q}{2\pi}\left[\ln\frac{2q}{k_{\rm IR}}-1\right]\\
		&-\frac{q^2}{4\pi k_{\rm UV}}
		+\mathcal{O}\!\left(\frac{q^4}{k_{\rm UV}^3},
		\frac{k_{\rm IR}^2}{q}\right).
	\end{split}
	\label{eq:app-p22-reg-asymptotic}
\end{equation}
The linear ultraviolet and logarithmic infrared terms are thereby separated explicitly.

\subsection{Vacuum subtraction and the physical transverse integral}

As in the number equation, we subtract $1/2$ per momentum mode in each real-field channel with identical integration limits. For the transverse channel, we define
\begin{eqnarray}
	P_{22}^{\rm IR}
	&\equiv&\lim_{k_{\rm UV}\to\infty}
	\left[P_{22}^{\rm reg}
	-\frac{k_{\rm UV}-k_{\rm IR}}{2\pi}\right]\nonumber\\
	&=&\frac{1}{2\pi}\int_{k_{\rm IR}}^\infty dk
	\left(\frac{\sqrt{k^2+q^2}}{k}-1\right).
	\label{eq:P22}
\end{eqnarray}
The subtracted kernel decays as $q^2/(2k^2)$, which permits
the limit $k_{\rm UV}\to\infty$. The subtraction sets the
equal-time vacuum reference according to the normal-ordered
particle density. The coupling $g_0$ retains its definition
as the physical 1D contact coupling~\cite{Pitaevskii2016}.

Subtraction of $k$ from the primitive in Eq.~\eqref{eq:app-p22-primitive} yields a function that vanishes as $k\to\infty$. Therefore,
\begin{equation}
	\begin{split}
		P_{22}^{\rm IR}
		=\frac{1}{2\pi}\biggl[
		&k_{\rm IR}-\sqrt{k_{\rm IR}^2+q^2}\\
		&+q\ln\frac{q+\sqrt{k_{\rm IR}^2+q^2}}{k_{\rm IR}}
		\biggr],
	\end{split}
	\label{eq:app-p22-sub-exact}
\end{equation}
consistent with Eq.~\eqref{eq:p22-ir-exact}. Expanding the two functions near the lower endpoint,
\begin{eqnarray}
	\sqrt{k_{\rm IR}^2+q^2}
	&=&q+\frac{k_{\rm IR}^2}{2q}
	+\mathcal{O}(k_{\rm IR}^4/q^3),\\
	\ln\frac{q+\sqrt{k_{\rm IR}^2+q^2}}{k_{\rm IR}}
	&=&\ln\frac{2q}{k_{\rm IR}}
	+\frac{k_{\rm IR}^2}{4q^2}
	+\mathcal{O}(k_{\rm IR}^4/q^4),\nonumber\\
\end{eqnarray}
we obtain
\begin{equation}
	\begin{split}
		P_{22}^{\rm IR}
		={}&\frac{q}{2\pi}\left[\ln\frac{2q}{k_{\rm IR}}-1\right]
		+\frac{k_{\rm IR}}{2\pi}\\
		&-\frac{k_{\rm IR}^2}{8\pi q}
		+\mathcal{O}(k_{\rm IR}^4/q^3).
	\end{split}
	\label{eq:P221D}
\end{equation}
In terms of the dressed scale $M$, the leading result therefore reads
\begin{equation}
	\begin{split}
		P_{22}^{\rm IR}
		={}&\frac{\sqrt m\,M}{\sqrt2\pi\hbar}
		\left[\ln\frac{2\sqrt{2m}M}{\hbar k_{\rm IR}}-1\right]\\
		&+\frac{k_{\rm IR}}{2\pi}
		+\mathcal{O}\!\left(\frac{\hbar k_{\rm IR}^2}{\sqrt m\,M}\right).
	\end{split}
	\label{eq:app-p22-M}
\end{equation}
The logarithm must be retained: $k_{\rm IR}\ll q$ makes its argument large, so an expansion about unity changes the infrared result. The ultraviolet subtraction leaves its coefficient unchanged.

\subsection{Common-cutoff depletion and infrared interpretation}

To combine the longitudinal result in Eq.~\eqref{P11} with the transverse result in Eq.~\eqref{eq:app-p22-sub-exact}, both channels must employ identical infrared cutoffs and vacuum subtractions. The noncondensed density relative to the specified background is then the
combination
\begin{eqnarray}
	\rho_{\rm ex}^{\rm IR}
	&=&\frac12\left(P_{11}^{\rm IR}+P_{22}^{\rm IR}\right)\nonumber\\
	&=&\frac{1}{4\pi}\int_{k_{\rm IR}}^\infty dk
	\left(\frac{k}{\sqrt{k^2+q^2}}
	+\frac{\sqrt{k^2+q^2}}{k}-2\right).\nonumber\\
	\label{eq:app-depletion-integral}
\end{eqnarray}
The kernel is nonnegative because it has the form $x+x^{-1}-2$ with $x>0$. At large momentum it behaves as $q^4/(4k^4)$, so the depletion is ultraviolet convergent. Integration gives
\begin{equation}
	\begin{split}
		\rho_{\rm ex}^{\rm IR}=\frac{1}{4\pi}\biggl[
		&q\ln\frac{q+\sqrt{k_{\rm IR}^2+q^2}}{k_{\rm IR}}\\
		&-2\sqrt{k_{\rm IR}^2+q^2}+2k_{\rm IR}\biggr],
	\end{split}
	\label{eq:app-depletion-exact}
\end{equation}
and, for $k_{\rm IR}\ll q$,
\begin{equation}
	\rho_{\rm ex}^{\rm IR}
	=\frac{q}{4\pi}\left[\ln\frac{2q}{k_{\rm IR}}-2\right]
	+\frac{k_{\rm IR}}{2\pi}
	+\mathcal{O}(k_{\rm IR}^2/q).
	\label{eq:app-depletion-log}
\end{equation}
Although vacuum subtraction eliminates the ultraviolet divergence, the infrared dependence stays physical. In a periodic system of length
$L$, the smallest nonzero momentum is $2\pi/L$. The corresponding finite-size correlator is then a discrete sum,
\begin{equation}
	P_{22,L}^{\rm sub}
	=\frac{1}{2L}\sum_{n\neq0}
	\left[\sqrt{1+\frac{q^2}{(2\pi n/L)^2}}-1\right].
	\label{eq:app-p22-discrete}
\end{equation}
Choosing $k_{\rm IR}\sim2\pi/L$ reproduces the leading logarithm
in the continuum approximation. Matching to the discrete sum
fixes the finite additive constant. The constants in
Eqs.~\eqref{eq:P221D} and \eqref{eq:app-depletion-log} correspond
to the sharp continuum cutoff. A local calculation may use
$k_{\rm IR}\sim\ell_{\rm coh}^{-1}$ for the coherent region
defined in Sec.~\ref{sec:ir}.

At fixed coupling and density, the logarithmic growth as
$k_{\rm IR}\to0$ reflects the loss of global phase coherence
and persists after the ultraviolet vacuum subtraction.
The perturbative result applies at weak coupling while
$\rho_{\rm ex}^{\rm IR}/\rho\ll1$. At larger length scales,
the correlations require a phase-based
description~\cite{Haldane1981PRL,MoraCastin2003,Rigol2011}.
In the bulk closure, $P_{22}$ cancels from the equations for
$M$ and $\mu$, keeping the EOS infrared finite.

%

\end{document}